\documentclass[prl,aps,twocolumn,superscriptaddress]{revtex4-1}
\usepackage{amssymb,amsfonts,amsmath,graphicx,color,times,mathtools}
 \pdfoutput=1
\usepackage{subfig}
\usepackage[normalem]{ulem}
\usepackage{hyperref}

\usepackage{mathrsfs}
\usepackage{dcolumn}
\usepackage{bm}
\usepackage{mathbbol}
\usepackage{color}

\def\tr{\mbox{tr}}

{\catcode`\|=\active 
  \gdef\Braket#1{\begingroup
\mathcode`\|32768\let|\BraVert\left<{#1}\right>\endgroup}}
\def\BraVert{\egroup\,\mid\,\bgroup}

\def\Tr{\mbox{Tr}}

\definecolor{john}{rgb}{0.3,0.3,0.8}

\definecolor{michele}{rgb}{0,0.5,0.9}

\allowdisplaybreaks

\begin{document}

\title{Thermodynamics of quantum information scrambling}

\author{Michele Campisi}
\email{michele.campisi@sns.it}
\affiliation{NEST, Scuola Normale Superiore \& Istituto Nanoscienze-CNR, I-56126 Pisa, Italy}

\author{John Goold}
\email{jgoold@ictp.it}
\affiliation{The Abdus Salam International Centre for Theoretical Physics (ICTP), Trieste, Italy}

\date{\today}

\begin{abstract}
Scrambling of quantum information can be conveniently quantified by so called out-of-time-order-correlators (OTOC's), i.e. correlators of the type $\langle W^{\dagger}_\tau V^{\dagger}W_\tau V\rangle $, whose measurements presents a formidable experimental challenge. 
Here we report on a method for the measurement of OTOC's based on the so-called two-point measurements scheme developed in the field of non-equilibrium quantum thermodynamics. The scheme is of broader applicability than methods employed in current experiments and also provides a clear-cut interpretation of quantum information scrambling in terms of non-equilibrium fluctuations of thermodynamic quantities such as work. Furthermore, we provide a numerical example on a spin chain which highlights the utility of our thermodynamic approach when understanding the differences between integrable and ergodic behavior. We also discuss connections to some recent experiments. 
\end{abstract}

\maketitle

\makeatletter

\begin{figure}[b]
\includegraphics[width=1\linewidth]{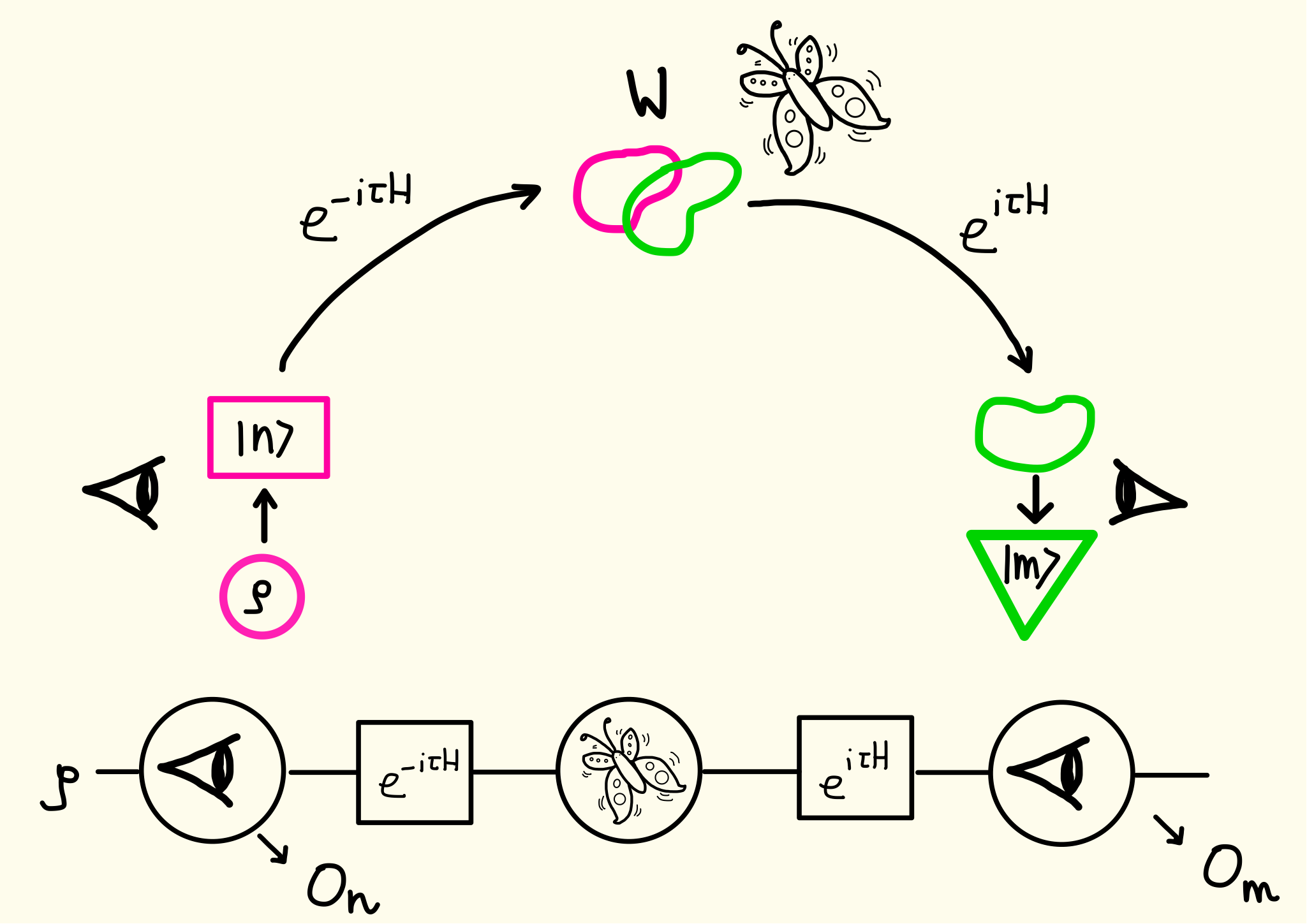}
 \caption{The Wing-flap protocol}
\label{fig:1}
\end{figure}
{\bf Introduction --}
Imagine preparing an omelette. You start from a couple of whole eggs and you scramble them, say by means of the rotary motion of a fork. Given the time-reversal symmetric microscopic laws of nature, you should be able to recover the whole eggs, if the inverse rotary fork motion is then applied. This in practice can never happen, because even the tiniest perturbation either in the backward fork motion, or in the microscopic state of the eggs, would prevent the eggs from un-scrambling, and would in fact result in further scrambling. Popular wisdom mirrors this truth with expressions like ``fare una frittata'' (``making an omelette'' in Italian) or ``making a hash'', meaning causing some situation for which there is no turning back. How macroscopic reversibility emerges from reversible microscopic dynamics is in fact one of the most studied problems in statistical mechanics since 
the early days of the famous debate between Boltzmann and Loschmidt \cite{Loschmidt:1876}.

Popular quantifiers of irreversibility are the Loschmidt echo \cite{gorin2006dynamics} and the irreversible entropy production \cite{spohn1978entropy}. Recently much attention has been devoted to yet another quantifier, namely the so called out-of-time-order correlator, namely correlator of the form:
\begin{equation}
\label{eq:otoc}
F_{V,W}(\tau)=\tr(\rho W^{\dagger}_\tau V^{\dagger}W_\tau V)
\end{equation}
When $W$ is a local unitary operator, the OTOC measures how much quantum information is scrambled during the dynamics \cite{Hayden:2007}.This is best understood by noticing that the OTOC can be directly related to the non-commutativity of $V$ and $W$ in time.\begin{equation}
C(\tau)=\langle [W_\tau ,V]^{\dagger}[W_\tau ,V]\rangle=2(1-\text{Re}[F_{V,W}(\tau)]).
\end{equation}
In the situation when $W$ is a local perturbation then $C(\tau)$ measures the growth and complexity of it's effect in time.
Furthermore an exponential time dependence of $F_{V,W}(\tau)$ reveals the presence of a butterfly effect and allows for extracting the corresponding Lyapunov exponent.
OTOC's were first introduced by Larkin and Ovchinnikov \cite{Larkin:1969} to investigate the validity of semi-classical approximations in the theory of superconductivity. They also appear in the context of the regression hypothesis, where due to their lack of time ordering they have been dubbed ``unartig Korrelatoren'' (i.e. ``naugthy correlators'') \cite{Talkner79thesis}.

Interest in OTOC's is currently undergoing a revival \cite{Hayden:2007,Sekino:2008,Shenker:2014,Maldacena:2015waa,Hosur:2016,Swingle:2016,Aleiner:2016,Swingle:2016jdj,Chen:2016mzk,adolfo,Halpern:2016,huse}, after Kitaev pointed out their relevance in the context of holographic duality \cite{kitaev_talk}. Accordingly a number of  experimental schemes have been recently proposed for the measurement of the OTOC and the first experimental measurements thereof have just been reported. Swingle et al. \cite{Swingle:2016} propose an interferometric scheme where the OTOC is encoded into the quantum state of an ancilla, to be implemented in a cold-atom set-up. 
Recent experiments, performed respectively with a trapped ion quantum magnet and NMR \cite{Garttner2016,Li2016} report a method for measuring infinite temperature (i.e., $\rho=\mathbb{1}$) OTOC's, i.e. OTOC's of the form
$
F_{V,W}(t)=\Tr(W^{\dagger}_t V^{\dagger}W_t V)
$.
The fact that $\rho$ does not appear here introduces a great simplification in the implementation of the scheme where now one can access $F_{V,W}(t)$ by preparing the system in the service state $\rho' \propto V+a\mathbb{1}$ ($\mathbb{1}$ stands for the identity operator) and measure the simpler 3-point correlator $\Tr(W^{\dagger}_t V^{\dagger}W_t \rho')$ instead
(see below for a more detailed discussion of this point).

Here we propose an alternative method to access $F_{V,W}(t)$ experimentally which does not require the employment of an ancillary system, nor of a service-state, and is applicable to any state $\rho$. The method is inspired by a scheme that proved extremely successful for the investigation of non-equilibrium quantum thermodynamics, namely the so called two-point measurement scheme \cite{Kurchan00arXiv,Tasaki00arXiv,Talkner07JPA40,Esposito09RMP81,Campisi11RMP83,Campisi11RMP83b,Hanggi15NATPHYS11}. Two point measurements schemes are protocols where an observable is measured twice, at the beginning and at the end of some non-equilibrium  manipulation carried on a system. The statistics of the observed change in the measured values encodes information about the manipulation. 
In our scheme an observable $O$ is projectively measured at the beginning and at the end of the information scrambling manipulation (which we dub the ``wing-flap protocol'' due to its connection with the idea of the Butterfly effect coming from non-linear physics), and the change of its value $\Delta O$ is recorded. 
In the wing-flap protocol a system evolves according to some forward evolution $e^{-i\tau H}$ and then goes back with the backward evolution $e^{i\tau H}$. The two evolutions are interrupted by a wing-flap (i.e. the application of some unitary operator $W$), that prevents the system to trace back to where it came from see Fig.~\ref{fig:1}. As we will see information about quantum information scrambling is encoded in the statistics of the observed changes in the measured values of $O$.

{\bf Main Results --} 
Let us express the unitary operator $V$ entering in Eq.~\eqref{eq:otoc} in exponential form:
\begin{align}
V &=e^{iuO} \label{eq:V-O}
\end{align}
with appropriate hermitian operator $O$ and real number $u$. Let us then consider the following protocol, which we dub the wing-flap protocol, see Fig. \ref{fig:1}:
\begin{enumerate}
\item Prepare the system in some state $\rho$.
\item Measure $O$.
\item Evolve the system with $H$ for a time $t=\tau$. 
\item Apply the wing-flap perturbation $W$. 
\item Evolve the system with $-H$ for a time $t=\tau$.
\item Measure $O$.
\end{enumerate}

The two measurements of $O$ are assumed to be projective giving eigenvalues $O_n$, $O_m$ respectively, and collapsing the system on the according eigenstates $|n\rangle, |m \rangle$. For simplicity we assume, without lack of generality, that the eigenvalues of $O$ are non-degenerate. We shall call the evolution under $H$, the forward evolution, and the evolution under $-H$, the backward evolution \footnote{The 5th step of the protocol can be replaced by (i) apply the anti-unitary time reversal operator $\Theta$ \cite{Messiah62Book} (that changes the sign of all momenta, and keeps positions unaltered)
(ii) Let the system evolve for a time $t=\tau$ under $H$ (iii) apply the anti-unitary time reversal operator $\Theta$ \cite{Messiah62Book}.
	The overall effect is $\Theta^\dagger e^{-i\tau H} \Theta = e^{i \tau H} $.
	If a magnetic field $B$ is present in the forward evolution, i.e., $H=H(B)$, one should reverse it in the backward evolution, that is in point (ii) above the evolution should occur under $H(-B)$ \cite{Campisi11RMP83}.
	}.

In absence of wing-flap, i.e. when $W $ is the identity operator $\mathbb{1}$, the state of the system at the end of the backward evolution would be exactly the initial state $|n\rangle$ of the forward evolution, and the second measurement would not alter it. When there is wing-flap $W \neq \mathbb{1}$, the state at the end of the backward evolution is in general a linear combination of all eigenstates of $O$, and the second measurement selects one of them, $|m\rangle$, which may differ from $|n\rangle$, due to the information scrambling.

On repeating the wing-flap protocol many times, in each realization the observed eigenvalues $O_n,O_m$ assume random values, and by repeating the protocol an infinite number of times one can build the probability density function (pdf) $p(\Delta O,\tau)$ of observing a change  
\begin{align}
\Delta O = O_m-O_n
\end{align}
in the quantity $O$. 
The statistics $p(\Delta O,\tau)$ contains information about the OTOC, Eq. (\ref{eq:otoc}). This connection can be established at a formal level as follows. Let 
\begin{align}
G(k,\tau) = \int p(\Delta O,\tau) e^{-i k \Delta O} d\Delta O
\label{eq:G-def}
\end{align}
be the characteristic function of $p(\Delta O,\tau)$, i.e., its Fourier transform \cite{Talkner07PRE75}. Our first main result is
\begin{align}
F_{e^{iuO},W}(\tau)  = G(u,\tau)
\label{eq:main}
\end{align}
Eq.~(\ref{eq:main}) says that the OTOC associated to the operators $W,e^{iu O}$ is identical to the characteristic function of the random variable $\Delta O$ in the wing-flap protocol above.

To prove the statement consider the formal expression of $p(\Delta O,\tau)$
\begin{align}
p(\Delta O,\tau) = \sum_{n,m} \delta[ \Delta O-O_m+O_n] P_\tau[m|n] p_n
\label{eq:p}
\end{align}
where $p_n= \Tr \, \Pi_n \rho$  is th probability of observing $O_n$ in the first measurement and $\delta(x)$ stands for Dirac's delta function. For simplicity, in the following we shall restrict to the case  $[\rho,O]=0$.
The symbol $P_\tau[m|n] $ stands for the probability of observing $O_m$ in the second measurement given that $O_n$ was observed in the first measurement:
\begin{align}
P_\tau[m|n] = |\langle m | U_\tau | n\rangle |^2
\label{eq:Pmn}
\end{align}
where $U_\tau$ is the unitary describing the evolution between the two measurements. It reads 
\begin{align}
U_\tau = e^{i\tau H} W e^{-i\tau H}
\label{eq:Utau}
\end{align}
Note that $U_\tau$ is the operator $W$ at time $\tau$ in the Heisenberg representation: $U_\tau=W_\tau$.
Using (\ref{eq:G-def},\ref{eq:p},\ref{eq:Pmn},\ref{eq:Utau}) and the resolution of identity $\sum_m |m\rangle\langle m |= \mathbb{1}$ and $[\rho,O]=0$, we obtain
\begin{align}
G(u,\tau) &= \sum_{n,m} e^{-i u O_m} e^{i u O_n}P_\tau[m|n] p_n \nonumber \\
&= \sum_{n,m}  \langle n | U^\dagger (\tau) | m\rangle e^{-i u O_m}  \langle m | U_\tau | n\rangle e^{i u O_n} p_n \nonumber \\
&= \sum_{n}  \langle n | U^\dagger_\tau  e^{-i u O}   U_\tau  e^{i u O} | n\rangle p_n \nonumber \\
&= \Tr \,  U^\dagger_\tau  e^{-i u O}   U_\tau  e^{i u O} \rho \nonumber \\
&= \Tr \, W^{\dagger}_{\tau} V^\dagger W_\tau V \rho 
 \end{align}
Thus the characteristic function of the statistics $p(\Delta O,\tau) $ of changes of $\Delta O$ is the OTOC between $e^{iu O}$ and $W$.

{\bf Thermodynamics of information scrambling --} The realization that OTOC's can be recast as two point measurement protocols allows us to directly connect with non-equilibrium quantum thermodynamics. Consider the case when the measured quantity $O$ is the Hamiltonian $H_0$ of a quantum system
$
O= H_0
$.
In this case Eq. (\ref{eq:main}) implies that $F_{e^{i u H_0},W}(\tau)$ is the characteristic function of work $G_\text{work}(u,\tau)$, namely the Fourier transform of the work probability distribution function
\begin{align}
p(w,\tau) = \sum_{n,m} \delta[ w -e_m+e_n] P_\tau[m|n] p_n
\end{align}
Here $w$ denotes work, i.e. the difference of final and initial measured system eigenenergies $e_k$.
In absence of the wing-flap no work is done on the system, whereas in general, when a wing-flap operator $W$ is present work is performed on the system, and the distribution drifts and spreads. 
The spread, namely the second moment of the work distribution $\langle w^2 \rangle = \int dw p(w,\tau) w^2= \Tr (W_\tau^\dagger H_0 W_\tau -H_0)^2\rho$ quantifies information scrambling as the expectation of the square of the commutator between $W_\tau$ and $H_0$
\begin{align}
\label{eq:w-variance}
\langle w^2 \rangle=  \langle [W_\tau,H_0]^\dagger [W_\tau,H_0] \rangle
\end{align}
In the case of an initial thermal equilibrium $\rho= e^{-\beta H_0}/Z$ ($\beta$ denotes the inverse temperature and $Z=\Tr e^{-\beta H_0}$ the partition function), the first moment $\langle w \rangle =\Tr (W_\tau^\dagger H_0 W_\tau -H_0)\rho$ of the work-distribution can be written in terms of the relative entropy $
S[\rho_\tau||\rho]$  between the initial state $\rho$, and its evolved $\rho_\tau$ \cite{Kawai07PRL98,Deffner10PRL105}:
\begin{align}
\label{eq:dissipation}
\langle w \rangle = \beta^{-1} S[\rho_\tau||\rho] =  \beta^{-1} \Tr (\rho_\tau \ln \rho_\tau - \rho_\tau \ln \rho)
\end{align}
The quantum relative entropy is a measure of the distinguishability of the two quantum states $\rho,\rho_\tau$ and its non-negativity is a consequence of the second law of thermodynamics \cite{Schloegl66ZP191}. It accordingly provides a meaningful quantifier of information scrambling occurring as a consequence of wing-flapping \footnote{In itself $S[\rho_\tau||\rho]$ is not a metric but it is known to bound the trace distance  via Pinsker's inequality: $S[\rho_\tau||\rho] \geq |\rho_{\tau} - \rho |_1^2/2$ \cite{Audenart:2014}}.

The characteristic function of work, the second moment of the work distribution, and the first moment all encode various aspects of information scrambling. We remark that all quantities, $F_{e^{iuH_0},W}(\tau)$, $ \langle |[W_\tau,H_0]|^2 \rangle
$ and $S[\rho_\tau||\rho]$, can be inferred from the work pdf $p(w,\tau)$. It is worth stressing that, if after the wing-flap protocol the system is brought back in contact with a thermal bath as to re-establish its initial thermal state, the quantity $\beta^{-1} S[\rho_\tau||\rho]$ is equal to the average heat $\langle q \rangle$ that the system releases in to the bath \cite{Apollaro15PSCT165}. Thus information scrambling might be accessed not only through work measurements, but also through heat measurements. Calorimetric measurements schemes being developed for low temperature solid state devices \cite{Solinas10PRB82,Gasparinetti15PRAPP3}, could be used for this purpose.

{\bf Illustrative example --}
\begin{figure*}
\includegraphics[width=1\linewidth]{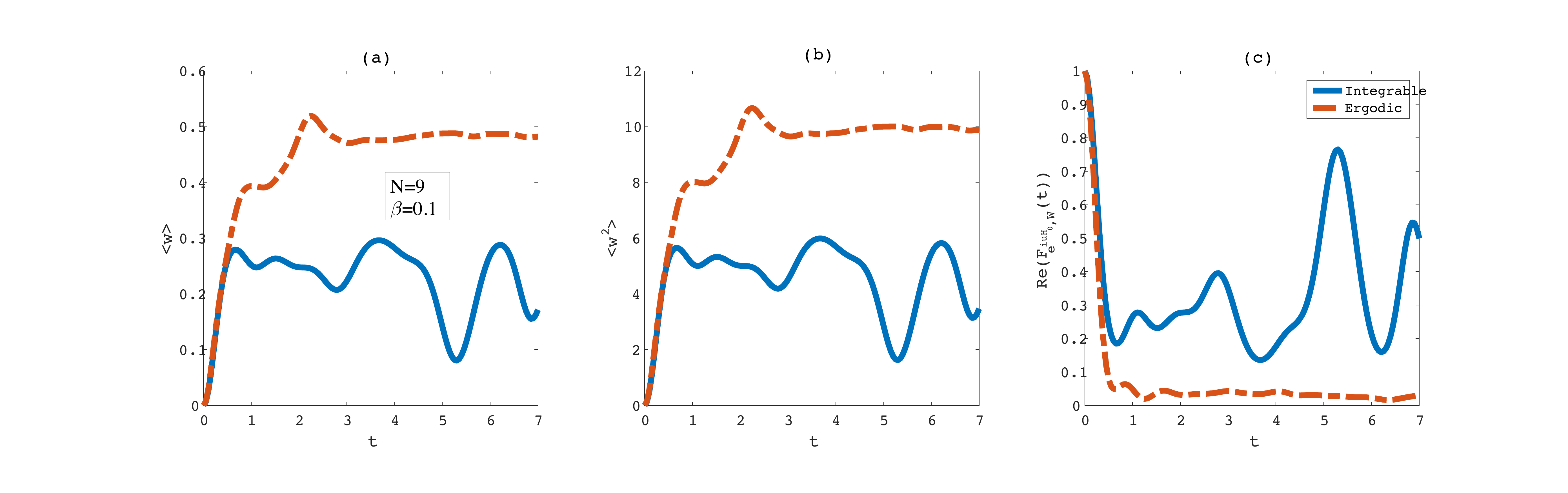}
\caption{Shown here are (a) the time dependent mean and (b) second moment of the work distribution for the quench protocol described in the text. (c) shows the real part of the OTOC $F_{e^{iuO},W}(\tau)$, for $u=1$.  Here we focus on a chain of $9$ spins at inverse temperature $\beta=0.1$ contrasting both integrable and non-integrable dynamics. We use the model and parameters specified in \cite{kim2013ballistic} ($J=1,g=0.90450849,h=0.8090169$)}
\label{fig:2}
\end{figure*}
In order to illustrate our results, we consider a spin chain of length $L$ described by the Hamiltonian
\begin{align}
\label{eq:quench1}
H_0 = g \sum_{i=1}^{L} \sigma_x^i
\end{align}
prepared in the thermal state
$
\rho = e^{-\beta H_0}/Z.
$
At time $t=0$ we turn on a perturbation $H_i$ so that the system evolves with the Hamiltonian $H=H_0+H_i$. At time $t=\tau/2$ we instantaneously apply the wing flap operator (a rotation about the x axis) $W=e^{-i\theta\sigma^k_x}$ with $\theta=\pi/2$. The system then evolves with $-H$ until time $\tau$. 
We consider the following forms of evolutions generated by Hamiltonians $H_i$:
\begin{align}
\label{eq:quench2}
H_\text{1} &=  J\sum_{i=1}^{L-1} \sigma_z^i\sigma_z^{i+1} \\
H_\text{2} &= J\sum_{i=1}^{L-1} \sigma_z^i\sigma_z^{i+1} + h \sum_{i=1}^{L-1} \sigma_z^i + (h-J) (\sigma_z^1+\sigma_z^L)
\end{align}
corresponding respectively to integrable and non-integrable dynamics (ergodic). This model was recently used by Kim and Huse in order to understand the phenomenology of entanglement growth in ergodic and integrable systems \cite{kim2013ballistic}. We choose identical parameters here.
Fig.~\ref{fig:2} shows the temporal behavior of the real part of the OTOC $F_{e^{iuH_{0}},W}(\tau)$, the mean and second moment of the associated work distributions for 9 spins. In all cases the wingflap operator acting on the central spin ($k=5$) and both integrable and ergodic evolutions are displayed.  Note how, as also shown in a closely related experiment \cite{Li2016}, that the integrable case is characterized by oscillations, i.e. recurrences, while no recurrence is observed in the time span over which the simulation is carried, in the ergodic case. This reflects the irreversible information scrambling occurring in the non-integrable case. Note also that, despite non-integrability, no exponential behavior is observed in the time dependence of the OTOC, this is in agreement with previous numerical \cite{Fine14PRE80} and experimental findings \cite{Li2016}. The plots reveal a small mean work as compared to the second moment $ \langle w \rangle^2 \ll \langle w^2 \rangle $ and a proportionality between them in accordance with linear response theory $\langle w \rangle \simeq \beta (\langle w^2 \rangle - \langle w \rangle^2)/2 \simeq \beta \langle w^2 \rangle/2 $ \cite{Hermans91JCP95,Jarzynski97PRL78}.

{\bf Interpreting existing experiments --} The two point measurement scheme allows for a straightforward interpretation of existing experiments, e.g. \cite{Li2016,Garttner2016}. These experiments, as mentioned in the introduction, measure infinite temperature OTOC's, by means of a service state $\rho'$. Let us now comment on the experiment of Ref.~\cite{Garttner2016}. There $V$ is proportional to $S_x=\sum_i \sigma_x^i$ that is the $x$ component of the total magnetization of a system of $N$ spins. The system is prepared in the factorized service-state $\rho'$ with all spin pointing up in the $x$ direction. It then evolves under some wing-flap unitary $U_\tau$ which we need not specify here. The authors measure  the expectation of $S_x$ at the end of the protocol. Interpreting $S_x$ as the initial Hamiltonian, i.e., $H_0 = S_x$, that is in fact a measurement of average work $\langle w \rangle $. Due to the special preparation $\rho'=H_0+(N/2)\mathbb{1}$, the latter assumes the form of an infinite temperature OTOC:
$
\langle w \rangle = \Tr (U^\dagger_\tau H_0 U_\tau -H_0)\rho' 
=\Tr U^\dagger_\tau H_0 U_\tau H_0 - \Tr H_0^2. 
$
The same measurement of $\langle w \rangle $ carried on a generic finite temperature thermal state $\rho \propto e^{-\beta H_0}$ would give the relative entropy $S[\rho_\tau||\rho]$ . Under the provision of linear response theory that would also give directly the second moment of the work distribution, hence the average square commutator $\langle |[W_\tau,H_0]|^2 \rangle$. These quantities can thus be straightforwardly accessed with the current measurement scheme, while the OTOC $F_{e^{iuO},W}(\tau)$ requires the implementation of the two-point measurement scheme.

{\bf Conclusions --} In summary we have put forward a new method for measuring information scrambling. The new method consists in a two-point projective measurement scheme. At variance with previously proposed schemes, the present one does not require the employment of an ancillary system nor the preparation of a service-state and is applicable to generic finite temperature conditions. According its scope of applicability is broader than current methods. The scheme not only offer a practical alternative for the measurement of OTOC's as compared to existing proposals, but also reveals a fundamental connection between non-equilibrium fluctuations of fundamental thermodynamics quantities, namely work and heat, and scrambling of information. The latter allows us to understand and interpret existing experiments in thermodynamic terms.

{\bf Acknowledgements --} M.C. was supported by Unicredit bank and by a Marie Curie Intra European Fellowship within the 7th European Community Framework Programme through the project NeQuFlux grant n. 623085. J.G. and M.C. are partially supported by the COST action MP1209 ``Thermodynamics in the quantum regime''. J.G. first heard of the topic of OTOC at Kitaev's lecture for the presentation of the 2015 Dirac Medal at ICTP \footnote{The ceremony can be seen at https://youtu.be/63ptOKbEECY} and would like to thank V.~Varma for convincing him that it was interesting. JG thanks A.~Silva and G.~Guarnieri for useful comments.

\bibliography{butterfly}

\end{document}